\documentclass{aastex701}
\usepackage{amsmath}
\usepackage{amssymb}

\begin{document}

\title{Making Euclid VIS Imaging AI-Ready: A Scalable Pipeline for Morphology, Anomaly Detection, and Similarity Search}

\author[
  orcid=0009-0008-0391-4437,
  gname={Jinhui},
  sname={Xie}
]{Jinhui Xie}
\affiliation{
National Astronomical Observatories,
Chinese Academy of Sciences,
Beijing 100101,
China
}
\affiliation{
University of Chinese Academy of Sciences, Beijing 100049, China
}
\affiliation{
National Astronomical Data Center,
Beijing 100101, China
}
\email{xiejinhui22@mails.ucas.ac.cn}

\author[orcid=0000-0002-7397-811X,gname=YUNFEI, sname='XU']{Yunfei XU} 
\affiliation{
National Astronomical Observatories,
Chinese Academy of Sciences,
Beijing 100101,
China
}
\affiliation{
National Astronomical Data Center,
Beijing 100101,
China
}
\email[show]{xuyf@nao.cas.cn}
\correspondingauthor{Yunfei Xu, Chenzhou Cui}

\author[gname=ZHEN, sname='ZHANG']{ZHEN ZHANG} 
\affiliation{
National Astronomical Observatories,
Chinese Academy of Sciences,
Beijing 100101,
China
}
\affiliation{
University of Chinese Academy of Sciences, Beijing 100049, China
}
\affiliation{
National Astronomical Data Center,
Beijing 100101, China
}
\email{zhangzhen@nao.cas.cn}

\author[gname=Lang, sname='CHEN']{Lang CHEN} 
\affiliation{
National Astronomical Observatories,
Chinese Academy of Sciences,
Beijing 100101,
China
}
\affiliation{
University of Chinese Academy of Sciences, Beijing 100049, China
}
\affiliation{
National Astronomical Data Center,
Beijing 100101, China
}
\email{chenlang@bao.ac.cn}

\author[orcid=0000-0002-7456-1826,gname=Chenzhou,sname=Cui]{Chenzhou CUI}
\affiliation{
National Astronomical Observatories,
Chinese Academy of Sciences,
Beijing 100101,
China
}
\affiliation{
National Astronomical Data Center,
Beijing 100101,
China
}
\email[show]{ccz@nao.cas.cn}

\begin{abstract}

The Euclid mission is delivering an unprecedented volume of high-resolution galaxy imaging, posing new challenges for standardized and reproducible reuse across morphological analyses. We present a scalable pipeline that converts released VIS cutouts into standardized $224\times224$ model inputs and uses a pretrained DINOv2 ViT-S/14 feature extractor to construct a 365,513-row, 384-dimensional cutout-and-embedding product from the Galaxy Zoo Euclid (Q1) catalogue. Initial production of the 365,513 released cutouts required approximately 48 hours, corresponding to an estimated end-to-end rate of $2.12$ cutouts s$^{-1}$. The service supports batch task management, while persistent reuse of generated products avoids repeating mosaic extraction for matching requests; together with released-scale production, these mechanisms define the operational scalability considered here. The representation is evaluated with held-out regression, few-label classification, anomaly-candidate prioritization, and similarity retrieval. Held-out ridge probes yield $R^2=0.377$--$0.596$ across four catalogue quantities. Under a one-percent total labelled budget that includes validation, frozen MLP macro-F1 is 0.802--0.861 for three tasks but 0.484 for the strongly imbalanced spiral task; limited final-block fine-tuning gives 0.810--0.850 for those three tasks and 0.520 for spiral. The historical anomaly workflow identifies 1,681 configuration-dependent candidates, and the displayed examples illustrate image-quality failures without estimating their prevalence. These results show that a standardized image interface and a reusable pretrained embedding can support several downstream analyses at the scale of the released Euclid Q1 sample, while the quantitative performance and interpretation remain task dependent under the declared input and evaluation contracts.

\end{abstract}



\section{Introduction} 
\label{sec:1}

The rapid growth of contemporary astronomical surveys has increased the need for data products that can be reused across machine-learning analyses. For imaging surveys, such reuse requires standardized pixel inputs, traceable metadata, and a stable interface between representations and downstream tasks, rather than a separate preprocessing path for every experiment. Survey-scale studies have shown that learned image representations can support semantic similarity search and lightweight supervised probes over large unlabeled collections \citep{stein_similarity_2021,stein_mining_2022}, while supervised representation learning can provide reusable features for several galaxy-morphology tasks \citep{walmsley_practical_2022}. We use ``AI-ready'' in this operational engineering sense, not as a claim that one representation is sufficient for every scientific inference.

The Euclid Quick Data Release 1 (Q1) provides a direct setting in which to examine this form of AI readiness. Euclid VIS supplies homogeneous, high-resolution optical imaging, and Q1 includes complementary morphology products based on S\'ersic fitting and visual classifications \citep{quilley_euclid_2025,2025arXiv250315310E}. Galaxy Zoo established large-scale citizen-science morphology through volunteer inspection \citep{lintott_galaxy_zoo_2008,willett_galaxy_zoo_2013}; the Euclid Q1 visual morphology catalogue extends that framework with Zoobot predictions trained on Euclid-specific volunteer annotations \citep{2025arXiv250315310E}. We use its released vote-fraction estimates to define task-specific high-consensus evaluation subsets.

Classical morphology analyses quantify galaxy light distributions through profile fits and nonparametric statistics, including concentration, asymmetry, smoothness, Gini, and $M_{20}$ \citep{conselice_relationship_2003,lotz_new_2004}. The Euclid Q1 S\'ersic catalogue provides a physically interpretable reference for this class of measurement \citep{quilley_euclid_2025}. These measurements and the detailed Euclid visual-morphology catalogue \citep{2025arXiv250315310E} are central scientific products, but each represents a predefined set of quantities or labels. A reusable pixel representation is therefore complementary: it can expose one common feature interface to several downstream analyses while leaving task-specific validation and interpretation explicit.

Recent self-supervised and foundation-model approaches offer one route to such reusable features. DINOv2 provides a direct primary reference for transferable visual features learned without task labels \citep{2023arXiv230407193O}. In astronomy, self-supervised image representations have been evaluated for galaxy morphology \citep{hayat_self-supervised_2021,walmsley_practical_2022}, radio-continuum morphology \citep{lastufka_self-supervised_2024}, and survey-scale similarity search and rare-object classification \citep{stein_similarity_2021,stein_mining_2022}. AION-1 evaluates frozen encoders across morphology, retrieval, and property-estimation tasks and includes a frozen DINOv2 ViT-g/14 baseline with an MLP head for Galaxy Zoo 10 morphology \citep{parker_aion_2025}. That benchmark uses Legacy Survey images, a ten-class label contract, and a sample of approximately 8,000 galaxies, so its reported accuracy is not directly comparable to our Euclid Q1 binary tasks.

The closest Euclid Q1 comparisons delimit the novelty of the present work. Euclid Collaboration: Siudek et al. \citeyearpar{2025arXiv250315312E} trained the AstroPT multimodal foundation model on roughly 300,000 Euclid optical and infrared images and spectral energy distributions, and evaluated morphology classification, redshift estimation, similarity search, and outlier detection, including one-percent-label experiments. Euclid Collaboration: Walmsley et al. \citeyearpar{2025arXiv250315310E} released automated detailed morphology for approximately 378,000 bright or extended Q1 galaxies using Zoobot models fine-tuned on Euclid-specific volunteer annotations; those measurements supply the morphology probabilities used in our evaluation. We therefore do not claim priority for morphology classification, similarity search, outlier detection, or one-percent-label evaluation on Euclid Q1. Our narrower contribution is an independently released engineering and data path: a service that provides square $128\times128$ VIS cutouts, a documented conversion to standardized $224\times224$ model inputs, a public interface to fixed official-DINOv2 embeddings, one common downstream workflow, and held-out quantitative audits with explicit evidence boundaries. This workflow complements the Euclid-specific learned models and morphology catalogues rather than replacing them.

\begin{figure*}[ht!]
\plotone{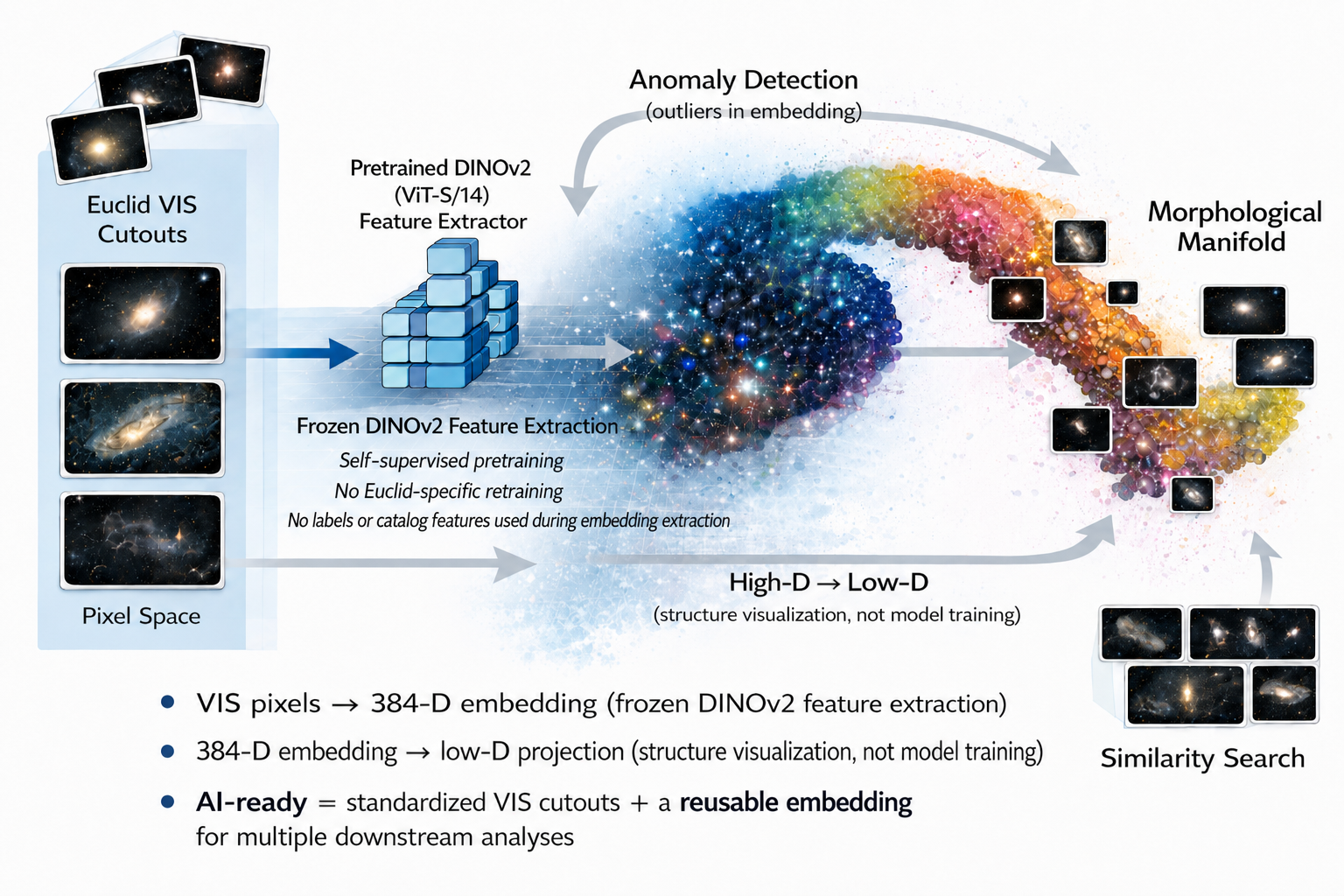}
\caption{Euclid VIS workflow for constructing an AI-ready representation. Released square $128\times128$ VIS cutouts are transformed into standardized $224\times224$ model inputs and processed by a pretrained DINOv2 ViT-S/14 feature extractor to produce 384-dimensional CLS-token embeddings. Here, AI-ready operationally denotes standardized, model-compatible image inputs together with a reusable embedding that supports regression, few-label classification, anomaly detection, and similarity retrieval within a common feature space.}
\label{fig:workflow}
\end{figure*}

In this work, we build a Euclid VIS pipeline that integrates standardized image preparation with a pretrained DINOv2 feature extractor. Galaxy cutouts are generated in bulk from the Euclid archive using a consistent preprocessing strategy and mapped to a 384-dimensional embedding without additional DINOv2 training. The operational meaning of ``AI-ready'' used throughout this paper is defined once in the caption of Figure~\ref{fig:workflow}. Low-dimensional projections are used for visualization, while held-out downstream evaluations quantify the information available in the original embedding.

We audit the fixed representation through held-out regression and few-label morphology classification, then retain anomaly ranking and similarity retrieval as exploratory demonstrations. The central goal is not to introduce a new foundation model or to establish priority for these downstream tasks, but to test whether a standardized public image-and-embedding interface can support them under clearly separated evidence contracts. Quantitative claims are restricted to held-out regression and classification results; the anomaly list is configuration dependent and dominated by quality-control cases, and the displayed retrieval examples remain qualitative.

\section{Data} \label{sec:2}

This section describes the data construction and preprocessing workflow used to transform Euclid VIS imaging into an AI-ready dataset. The emphasis is placed on combining large-scale survey imaging with morphology annotations and converting them into a standardized, model-compatible format suitable for downstream machine learning applications.

\subsection{Dataset} \label{subsec:2.1}

This work is based on imaging data from the Euclid Q1 and morphology annotations from the Galaxy Zoo project. Euclid VIS provides space-based optical imaging with a pixel scale of approximately $0.1''$ and a highly stable PSF, enabling consistent and high-resolution observations of galaxy structure over a wide survey area. These characteristics make Euclid VIS particularly well suited for morphology-oriented analysis.

Galaxy Zoo is a large-scale citizen science project in which galaxy morphologies are characterized through visual inspection by human volunteers \citep{lintott_galaxy_zoo_2008,willett_galaxy_zoo_2013}. Rather than assigning a single deterministic class to each object, Galaxy Zoo records answers along a decision tree of morphology questions \citep{walmsley_galaxy_2022}. The Euclid Q1 dynamic catalogue reports Zoobot predictions formatted as the fraction of volunteers expected to choose each answer when asked \citep{2025arXiv250315310E}. These model predictions are conditioned by the question tree and should not be interpreted as independent human labels for every object.

In this work, we adopt the Galaxy Zoo Euclid (Q1) catalogue, which combines Euclid VIS imaging with morphology measurements derived from Galaxy Zoo. This catalogue contains 380,111 unique objects \citep{walmsley_2025_15106473}. This number describes entries in the source morphology catalogue, whereas 365,513 is the number of VIS cutouts successfully produced and included in the released image dataset. The difference therefore separates catalogue entries from the released image-product rows; Table~\ref{tab:sample_flow} records these data-product boundaries explicitly. Removing nine duplicate embedding object IDs gives 365,504 unique catalogue--embedding matches for the principal analyses. The morphology measurements are generated by Zoobot models pretrained on Galaxy Zoo labels and fine-tuned on Euclid imaging. High-confidence Zoobot outputs agree strongly with held-out volunteer answers in the catalogue validation \citep{2025arXiv250315310E}, but they remain model predictions anchored to volunteer response distributions rather than an independent human-labelled test set.

We construct task-specific high-confidence labels from the catalogue outputs for controlled downstream evaluation. For each binary task, let $p_i^{+}$ and $p_i^{-}$ denote the catalogue fractions for the explicitly named positive and negative answers, and let $e_i$ indicate that the catalogue marks the question as applicable. The label is

\begin{equation}
y_{\mathrm{HQ},i}=
\left\{
\begin{array}{ll}
1, & e_i=1\ \mathrm{and}\ p_i^{+}>0.8, \\
0, & e_i=1\ \mathrm{and}\ p_i^{-}>0.8, \\
-1, & \mathrm{otherwise} .
\end{array}
\right.
\label{eq:hq_label}
\end{equation}

where $y_{\mathrm{HQ},i}=-1$ denotes an inapplicable or non-high-confidence case that is excluded from supervised evaluation. The dynamic catalogue predicts every answer internally but serializes child-question fractions as NaN when the expected leaf probability---the product of the answers leading to that question---is below 0.5 \citep{2025arXiv250315310E}. We use this finite/NaN leaf mask as the authoritative applicability indicator. Thus, edge-on requires the featured-or-disk branch, spiral requires featured-or-disk followed by not edge-on, and round/cigar requires the smooth branch. For the three-answer roundedness question, a negative label specifically requires \texttt{cigar-shaped}$>0.8$; a low round fraction is not sufficient because it may instead indicate the in-between answer.

Accordingly, the subset used for downstream classification is selected through the indicator
\begin{equation}
\mathcal{S}_{\mathrm{HQ}} = \left\{ i \; \middle| \; y_{\mathrm{HQ},i} \in \{0,1\} \right\},
\label{eq:hq_subset}
\end{equation}
so that only applicable galaxies with an explicitly high-confidence positive or negative answer are retained. The exact catalogue columns, parent paths, thresholds, class counts, split membership, and hashes are recorded in the released label contract.

Based on these high-confidence answers, we define four independent binary tasks: smooth versus featured, spiral versus non-spiral, edge-on versus non-edge-on, and round versus cigar. The first three use the corresponding paired answer columns; round versus cigar excludes the third, in-between answer. These cohorts evaluate agreement with high-confidence Zoobot outputs under the stated question-tree contract and are not presented as an independent validation against new human classifications.

This task-specific labeling strategy ensures that each classification problem is well-defined and minimally affected by ambiguity in other morphological dimensions, while enabling a direct evaluation of how the learned representation captures distinct aspects of galaxy structure. Since all tasks are derived from the same underlying vote-fraction space, the resulting labels remain anchored to the original human-informed morphology annotations. In this sense, our dataset is not a direct reuse of the released catalog, but a quality-controlled subset in which probabilistic morphology information is converted into high-confidence labels for controlled downstream evaluation within an AI-ready framework. The full data processing and analysis pipeline used in this work is publicly available at \url{https://github.com/xiejhhhhhh/Making-Euclid-VIS-Imaging-AI-Ready}.

\subsection{AI-ready Cutout Pipeline} \label{subsec:2.2}

A central component of this work is the construction of a standardized image dataset through a dedicated batch cutout pipeline. Euclid data products are not natively provided in a format optimized for pixel-level machine learning, and repeatable extraction of image cutouts is required to enable AI-ready analysis.

The underlying Euclid Q1 imaging data are hosted on the National Astronomical Data Center (NADC). The NADC is a national-level scientific data infrastructure in China, operated under the Chinese Academy of Sciences, and serves as a core platform for the collection, curation, and dissemination of astronomical data. It provides standardized data services, high-performance storage, and computing resources that support large-scale data-driven research in astronomy.

Based on this infrastructure, we develop a Euclid Image Cutout Service that enables automated extraction of galaxy-centered image stamps from the Euclid archive. The service is deployed on the NADC scientific computing platform and directly accesses the locally hosted Euclid Q1 mirror without a remote data-transfer step. The system operates on FITS-format catalogs containing celestial coordinates and supports batch processing of galaxy samples.

In practical use, the pipeline reads source positions from input catalogs and extracts corresponding image cutouts from the Euclid VIS data products. The service accepts catalog-based batch tasks, and independent requests can be executed in parallel across process workers. It scans persistent band directories for matching products using the target identifier, band, instrument, and file type; it also checks a local task cache keyed by coordinates, cutout size, instrument, file type, and band. Matching products are copied into the task output without repeating mosaic extraction, whereas unmatched requests are sent to the parallel cutout routine and newly generated products are retained for later reuse. This combination of batch task management, parallel handling, and persistent product reuse defines the operational scalability considered here. Based on the authors' operational record, initial production of the 365,513 released cutouts on the NADC deployment required approximately 48 hours, giving an estimated end-to-end rate of $2.12$ cutouts s$^{-1}$ (approximately 7,615 cutouts h$^{-1}$). The 48-hour duration is an author-attested operational estimate rather than a timestamp-complete benchmark. The rate describes initial release production, whereas persistent product reuse applies to subsequent matching requests; these mechanisms characterize different stages of the operational workflow.

The released VIS FITS cutouts used by the embedding extractor are square $128\times128$ arrays centered using catalogue astrometry. The historical model loader deterministically converts each array to the $224\times224$ DINOv2 input size as part of the pixel-to-tensor transform; the released FITS files themselves are not claimed to be stored at $224\times224$.

The historical transform is now stated operationally. Each image is read as \texttt{float32}; NumPy \texttt{nan\_to\_num} is applied; the 1st and 99th percentiles are computed per image; values are clipped to this interval and min--max scaled to $[0,1]$; the array is converted from $128\times128$ to $224\times224$ with the retained NumPy \texttt{resize} operation; and the grayscale array is replicated into three identical channels. No ImageNet or DINOv2 channel mean--standard-deviation normalization is applied. An evenly spaced audit of 10 input files found only $128\times128$ arrays and no non-finite pixels, while the complete $365{,}513\times384$ embedding array contained no non-finite values.

The cutout pipeline is implemented as a standalone, publicly available service for reproducible batch preprocessing. The implementation is released at \url{https://github.com/xiejhhhhhh/Euclid-Image-Cutout-Service} and provides a modular interface for batch cutout generation. While developed for Euclid Q1 data, the system can be adapted to other imaging surveys with similar data formats, subject to survey-specific validation.

By combining NADC data access with a dedicated batch cutout service, this pipeline transforms heterogeneous Euclid imaging data into a uniform dataset suitable for machine learning. The resulting dataset has been publicly released via the NADC and is accessible at \url{https://nadc.china-vo.org/res/r101833/}. It contains 365,513 released VIS image cutouts and is designed to support reproducible studies of galaxy morphology and representation learning. The verified row count and approximately 48-hour production record establish operation at the released data scale. Batch task management and persistent product reuse support subsequent matching requests without repeating mosaic extraction; the initial-production rate and reuse mechanism therefore describe complementary stages of the operational workflow.

Table~\ref{tab:sample_flow} reconciles the 380,111 unique catalogue objects, 365,513 released cutouts and embedding rows, the 365,504-object de-duplicated join, and the task-specific evaluation cohorts and class counts.

\begin{deluxetable*}{lrrrrrl}
\tabletypesize{\scriptsize}
\tablewidth{0pt}
\tablecaption{Sample flow, independently fixed object-ID evaluation cohorts, and classification class composition.\label{tab:sample_flow}}
\tablehead{\colhead{Cohort} & \colhead{N} & \colhead{Train N} & \colhead{Test N} & \colhead{Train $N_-/N_+$} & \colhead{Test $N_-/N_+$} & \colhead{Definition / exclusion}}
\startdata
high quality catalog & 380,111 & -- & -- & -- & -- & Unique source-catalogue object IDs \\
embedding rows & 365,513 & -- & -- & -- & -- & Released VIS cutout and embedding rows \\
embedding unique object ids & 365,504 & -- & -- & -- & -- & Nine duplicate embedding IDs removed \\
catalog embedding join & 365,504 & -- & -- & -- & -- & One-to-one matched analysis universe \\
regression ellipticity & 365,504 & 292,403 & 73,101 & -- & -- & Finite ellipticity; independently split \\
regression kron radius & 365,504 & 292,403 & 73,101 & -- & -- & Finite Kron radius; independently split \\
regression mu max & 365,494 & 292,395 & 73,099 & -- & -- & Finite peak surface brightness; independently split \\
regression mumax minus mag & 365,491 & 292,393 & 73,098 & -- & -- & Finite peak-minus-total magnitude; independently split \\
fewlabel smooth vs featured & 95,736 & 76,588 & 19,148 & 24,169/52,419 & 6,043/13,105 & High-confidence paired answers; root question \\
fewlabel spiral vs nonspiral & 50,538 & 40,430 & 10,108 & 1,257/39,173 & 314/9,794 & Featured and not-edge-on leaf mask \\
fewlabel edgeon vs nonedgeon & 109,193 & 87,354 & 21,839 & 74,611/12,743 & 18,653/3,186 & Featured leaf mask \\
fewlabel round vs cigar & 52,749 & 42,199 & 10,550 & 6,254/35,945 & 1,563/8,987 & Smooth leaf mask; in-between excluded \\
\enddata
\tablecomments{The catalogue contains 380,111 unique object IDs, whereas the released cutout/embedding product contains 365,513 rows. Removing nine duplicate embedding IDs leaves 365,504 matched objects. Each regression target and classification task is filtered and split independently with \texttt{random\_state}=42. For classification rows, $N_-/N_+$ denotes negative/positive counts after the paired high-confidence Galaxy Zoo question-tree applicability mask; the round-versus-cigar task excludes the in-between answer.}
\end{deluxetable*}

\section{Methods} \label{sec:3}

The methodological goal of this work is not to optimize a new morphology model for Euclid, but to construct an AI-ready representation pipeline that can support multiple downstream analyses within a unified framework. To this end, we adopt a pretrained vision transformer backbone, use it to generate embeddings for the released 365,513-cutout sample, and then evaluate the scientific utility of this representation through few-label classification, anomaly detection, and similarity-based retrieval.

\subsection{Representation Learning} \label{subsec:3.1}

The representation backbone adopted in this work is the official DINOv2 model \citep{2023arXiv230407193O}, used as a pretrained vision foundation model without additional retraining on Euclid data. We employ the \texttt{dinov2\_vits14} configuration, corresponding to a Vision Transformer (ViT-S/14) architecture with patch size $14\times14$ and embedding dimension $D=384$.

The DINOv2 model is pretrained on a curated dataset of approximately $1.42 \times 10^{8}$ images, selected from a much larger pool of web-scale data. We use this released checkpoint as a fixed reference representation for limited-label Euclid experiments. Its use here does not establish that generic pretraining is intrinsically superior to domain-specific self-supervised training; such a conclusion would require a matched architecture, input, split, and evaluation contract.

At the architectural level, the Vision Transformer processes an image by dividing it into non-overlapping patches. For an input image $\tilde{\mathbf{x}}_i \in \mathbb{R}^{3\times224\times224}$, the image is partitioned into a sequence of patches of size $14\times14$, resulting in $N= (224/14)^2 = 256$ patch tokens. Each patch is linearly projected into a $D$-dimensional embedding space and combined with a learnable positional encoding. A special classification token is prepended to the sequence, which aggregates global information through multi-head self-attention layers.

Formally, the transformer encoder can be written as
\begin{equation}
\mathbf{z}_i = f_{\theta}(\tilde{\mathbf{x}}_i),
\end{equation}
where $\mathbf{z}_i \in \mathbb{R}^{384}$ denotes the final CLS-token representation. The self-attention mechanism allows each token to attend to all others, enabling the model to capture both local structures (e.g., spiral arms) and global morphology (e.g., bulge--disk contrast) within a unified framework.

After the explicit per-image transform in Section~\ref{subsec:2.2}, each single-channel VIS array is replicated across three channels before being passed to the model,
\begin{equation}
\tilde{\mathbf{x}}_i = [\mathbf{x}_i, \mathbf{x}_i, \mathbf{x}_i].
\end{equation}
For the released lineage, no channel-wise mean--standard-deviation normalization is added after replication. All primary embeddings, UMAP projections, LOF rankings, retrieval examples, and main classification and regression results use this released transform. All embeddings are extracted in inference mode with frozen model parameters, and the normalized CLS token is retained as a 384-dimensional vector. No task-specific retraining or domain adaptation is performed during the representation-learning stage.

\subsection{Embedding Construction} \label{subsec:3.2}

Using the pretrained DINOv2 backbone described in Section~\ref{subsec:3.1}, we extract a fixed-dimensional representation $\mathbf{z}_i \in \mathbb{R}^{384}$ for each galaxy image through a forward pass in inference mode. A full-array audit found zero non-finite values and therefore removed or replaced no embedding rows or dimensions. For analyses that require feature scaling, the retained embeddings are standardized using
\begin{equation}
\tilde{\mathbf{z}}_i = \frac{\mathbf{z}_i - \boldsymbol{\mu}}{\boldsymbol{\sigma}},
\end{equation}
where $\boldsymbol{\mu}$ and $\boldsymbol{\sigma}$ are vectors estimated separately for each of the 384 feature dimensions on the relevant training partition. We use $\tilde{\mathbf{z}}$ for the held-out regression probes, with the scaler fitted on each target-specific training partition. The saved UMAP visualization and historical LOF calculation instead use the raw embedding $\mathbf{z}$, while retrieval uses the L2-normalized raw embedding.

For retrieval, given two raw embeddings $\mathbf{z}_i$ and $\mathbf{z}_j$, cosine similarity is defined as
\begin{equation}
s_{ij} = \frac{\mathbf{z}_i^\top \mathbf{z}_j}{\|\mathbf{z}_i\| \|\mathbf{z}_j\|}.
\end{equation}
This equation therefore uses $\mathbf{z}$ rather than the standardized $\tilde{\mathbf{z}}$ defined above; L2 normalization is applied immediately before evaluating the dot product.

To visualize local neighborhood structure in the representation, we use the saved UMAP projection generated from the raw embedding with $n_{\mathrm{neighbors}}=15$, $\mathrm{min\_dist}=0.1$, the default Euclidean metric, two output dimensions, and \texttt{random\_state}=42. The resulting mapping
\begin{equation}
\mathbf{u}_i = f_{\mathrm{UMAP}}(\mathbf{z}_i)
\end{equation}
provides a low-dimensional projection for descriptive visualization in Figure~\ref{fig:embedding_umap}; it is not evidence that global distances or a continuous physical manifold are preserved. The historical anomaly workflow described in Section~\ref{subsec:3.4} separately uses distance from the centroid of a two-dimensional UMAP projection as an explicit secondary filter. We distinguish that procedural use from the visualization role here.

We assess linearly recoverable catalogue information through ridge probes on four SourceExtractor-related fields: \texttt{ellipticity}$=1-B_{\mathrm{IMAGE}}/A_{\mathrm{IMAGE}}$ (dimensionless), \texttt{kron\_radius} (the major semi-axis of the elliptical Kron aperture, in pixels), \texttt{mu\_max} (peak surface brightness above the background, in mag arcsec$^{-2}$), and \texttt{mumax\_minus\_mag}$=\mathrm{MU\_MAX}-\mathrm{MAG\_AUTO}$ (mag). For each target variable $q_i$, we train a ridge regression model
\begin{equation}
\hat{q}_i = \mathbf{w}^\top \tilde{\mathbf{z}}_i + b,
\end{equation}
where the weight vector $\mathbf{w}$ is estimated by minimizing a regularized least-squares objective. Each target is filtered for finite values and then independently assigned to an 80\% training partition and a 20\% held-out test partition with NumPy seed 42; there is no single global assignment shared across all targets and tasks. The feature scaler is fitted on the target-specific training partition only. RidgeCV uses generalized cross-validation on that partition over $\lambda\in\{0.001,0.003,0.01,0.03,0.1,0.3,1,3,10,30,100,300,1000\}$ before predictions are generated for held-out objects. Membership hashes are included in the release contract.

Model performance is evaluated on the held-out partition using the coefficient of determination ($R^2$), mean absolute error (MAE), root-mean-square error (RMSE), and both Pearson and Spearman correlation coefficients. The training and test sample sizes and all five metrics are reported in Table~\ref{tab:regression}.

\subsection{Few-label Classification} \label{subsec:3.3}

To evaluate the effectiveness of the representation for supervised tasks, we perform a controlled few-label classification benchmark using the high-confidence morphology labels described in Section~\ref{subsec:2.1}. The objective is to quantify how classification performance scales with the fraction of labeled data.

The validated benchmark uses the official DINOv2 embedding and a task-specific fixed object-ID outer split. After applying the question-tree and high-confidence label contract, each task is independently stratified with \texttt{random\_state}=42 into an 80\% development partition and a 20\% held-out test partition; the same test objects are used for all heads and label fractions within that task. Label fractions of 1\%, 5\%, 10\%, and 100\% define the total labelled supervision budget sampled from the development partition. For each seed in $\{42,43,44\}$, that budget is then stratified 80/20 into training and validation subsets. Thus all labels used for epoch or checkpoint selection are included in the declared budget rather than supplied as an additional validation set.

Two classifier heads are evaluated: a linear probe and a shallow MLP with one 128-unit hidden layer, GELU activation, dropout 0.1, and a sigmoid decision threshold of 0.5. Frozen probes use AdamW with learning rate $10^{-3}$, weight decay $10^{-4}$, batch size 4096, and 25 epochs; binary cross-entropy uses a positive-class weight computed from the current training subset, and the checkpoint with the highest budget-internal validation macro-F1 is retained. The frozen-backbone experiment on the fixed 384-dimensional embeddings is reported in Section~\ref{subsec:4.2}. The separate 1\% limited-fine-tuning configuration uses one head-only warm-up epoch followed by training of the final transformer block, final normalization layer, and head for up to four epochs. It uses batches of 32, evaluation batches of 128, head and backbone learning rates of $10^{-3}$ and $10^{-5}$, respectively, weight decay $10^{-4}$, the same weighted loss and threshold, and budget-internal validation macro-F1 for checkpoint selection.

Macro-F1 is computed once on the fixed held-out test objects. Table~\ref{tab:fewlabel} reports the one-percent MLP comparison together with negative/positive recall, class composition, and trivial majority and stratified-random references; the complete linear/MLP and 1/5/10/100\% seed-level matrix is retained in the machine-readable release. Historical random-initialization and Euclid-specific self-supervised learning (SSL) records are not included in this matched benchmark because their retained runs do not share the complete architecture, input, label, and object-ID split contract. Published AION-1 and related external benchmark values are reported separately in the Results.

Historical Euclid-specific self-supervised learning (SSL) and random-initialization runs are documented separately in Appendix~\ref{app:historical_ssl}. The Euclid SSL checkpoint was produced by an eight-epoch DINO teacher--student run on the released Euclid VIS cutouts, using a local single-channel small-ViT implementation with 256-dimensional features. Because these runs do not match the official DINOv2 ViT-S/14 configuration in architecture, input channels, feature dimension, or retained split contract, they are excluded from the matched official-DINOv2 benchmark reported in Section~\ref{subsec:4.2} and Table~\ref{tab:fewlabel}.

\subsection{LOF-based Anomaly Detection} \label{subsec:3.4}

An AI-ready representation should support not only classification but also data-quality control and discovery-oriented exploration. To test this capability, we perform anomaly detection directly in the learned embedding space using the Local Outlier Factor (LOF) algorithm introduced by \citet{breunig_lof_2000}; \citet{baron_perez_classification_2025} provides recent astronomy context for self-supervised radio-source analysis. Let $\mathbf{z}_i \in \mathbb{R}^{D}$ denote the high-dimensional embedding of source $i$. For a neighborhood size $k$, the LOF score is defined as
\begin{equation}
\mathrm{LOF}_{k}(i)=\frac{1}{|N_k(i)|}\sum_{j\in N_k(i)}\frac{\mathrm{lrd}_{k}(j)}{\mathrm{lrd}_{k}(i)},
\end{equation}
where $N_k(i)$ is the set of $k$ nearest neighbors of source $i$ and $\mathrm{lrd}_{k}(i)$ is its local reachability density. Sources with elevated LOF values occupy locally underdense regions and are treated as candidates for inspection rather than confirmed astrophysical anomalies.

In the historical DINOv2 analysis used for the anomaly analyses reported in Section~\ref{subsec:4.3}, LOF is applied to the original, unstandardized 384-dimensional embeddings with the scikit-learn default Euclidean metric, $k=100$, \texttt{contamination}=0.1, and \texttt{novelty=False}. The negative outlier factor is converted into a positive score for visualization and ranking. The workflow also reuses the same saved two-dimensional UMAP projection described in Section~\ref{subsec:3.2} ($n_{\mathrm{neighbors}}=15$, $\mathrm{min\_dist}=0.1$, Euclidean metric, and \texttt{random\_state}=42), retaining objects beyond the 98th percentile of distance from that projection's centroid. The final candidate subset is the intersection
\begin{equation}
\mathcal{O}_{\mathrm{final}}=\mathcal{O}_{\mathrm{LOF}}\cap \mathcal{O}_{\mathrm{UMAP}},
\end{equation}
which selects objects that are both locally sparse in the original embedding and distant from the centroid of this particular UMAP projection. Because UMAP does not preserve global distances and both stages depend on hyperparameters, the resulting 1,681 objects are an exploratory, configuration-dependent candidate list. We use it primarily to demonstrate quality-control triage and candidate prioritization, not as a statistically complete catalog of astrophysical anomalies.

\subsection{Similarity-based Retrieval} \label{subsec:3.5}

The final component of the workflow is morphology-aware similarity retrieval. Given a query source $q$ with embedding $\mathbf{z}_q$, candidate neighbors are first identified in the original high-dimensional embedding space through cosine similarity,
\begin{equation}
s_{\mathrm{emb}}(q,i)=\frac{\mathbf{z}_q\cdot\mathbf{z}_i}{\|\mathbf{z}_q\|\,\|\mathbf{z}_i\|}.
\end{equation}
In the implementation, all embeddings are normalized before retrieval, so the cosine score is equivalent to the dot product between normalized vectors. Retrieval is performed directly in the embedding space rather than in the UMAP projection, with a candidate pool of 300 objects and a final return list of the top 4 matches.

On top of this embedding-based candidate selection, we introduce a lightweight re-ranking step to improve robustness against observational artifacts and ambiguous cases. The final retrieval score is defined as
\begin{equation}
S(q, i) = w_{\mathrm{emb}}\, s_{\mathrm{emb}}(q, i) + w_{\mathrm{lab}}\, s_{\mathrm{lab}}(q, i) + w_{\mathrm{phys}}\, s_{\mathrm{phys}}(q, i) + w_{\mathrm{qual}}\, s_{\mathrm{qual}}(i),
\end{equation}
where $s_{\mathrm{emb}}$ is the cosine similarity in embedding space, and the additional terms incorporate weak priors based on morphology labels, physical parameters, and image quality indicators.

The task-dependent weights used for the historical examples reported in Section~\ref{subsec:4.4} were manually specified heuristics rather than learned parameters. Spiral and non-spiral retrieval used $(w_{\mathrm{emb}},w_{\mathrm{lab}},w_{\mathrm{phys}},w_{\mathrm{qual}})=(0.70,0.22,0.06,0.02)$, while round, cigar, edge-on, and non-edge-on retrieval used $(0.62,0.24,0.12,0.02)$. We therefore treat these panels as qualitative demonstrations and do not interpret the displayed combined score as a calibrated probability.

As a limited sensitivity check on a fixed held-out query set, we also compare embedding-only retrieval with predeclared embedding/physical mixtures of 0.7/0.3 and 0.5/0.5. The weights are fixed before query labels or retrieval scores are loaded, and label information is used only for evaluation. This is a two-component ablation of the embedding and physical terms, not a direct validation of the four-component task-dependent weights used in the historical figures. The four-query-per-task check is therefore used to demonstrate a transparent held-out evaluation protocol, not to establish a generally optimal weighting.

\section{Results} \label{sec:4}

\subsection{Embedding Projections and Regression Probes} \label{subsec:4.1}

\begin{figure*}[ht!]
\plotone{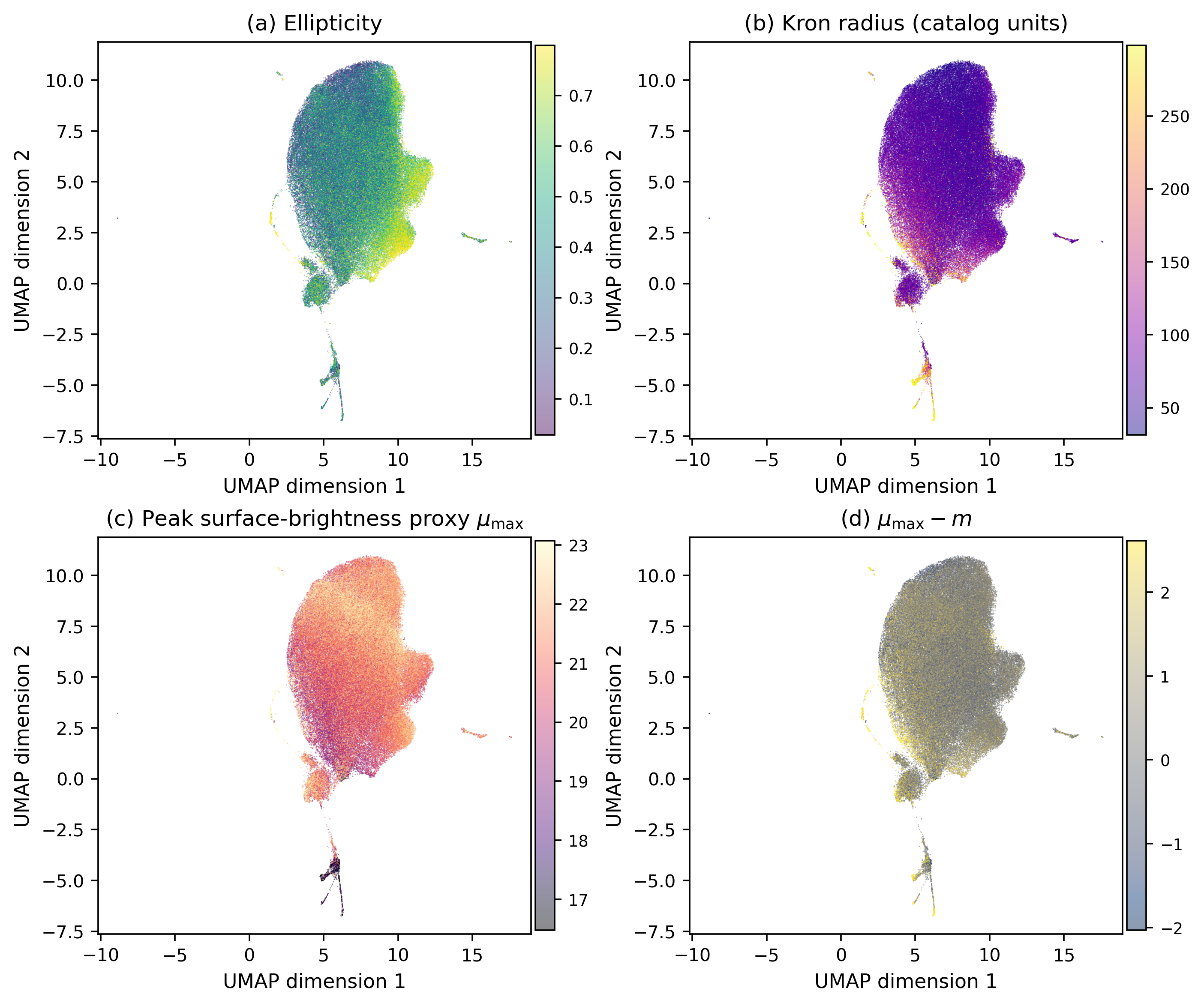}
\caption{Two-dimensional UMAP views of the fixed DINOv2 embedding. Panels show (a) ellipticity, (b) Kron radius, (c) $\mu_{\max}$, and (d) $\mu_{\max}-m$ for a 120,000-object display sample selected with pandas \texttt{random\_state}=42. The saved projection was fitted to the raw embedding with $n_{\mathrm{neighbors}}=15$, $\mathrm{min\_dist}=0.1$, Euclidean distance, and UMAP \texttt{random\_state}=42. Colors are clipped to the 1st--99th percentiles for visualization. UMAP is used descriptively and does not preserve global distances.}
\label{fig:embedding_umap}
\end{figure*}

Figure~\ref{fig:embedding_umap} shows two-dimensional UMAP projections of the fixed DINOv2 embeddings colored by ellipticity, Kron radius, $\mu_{\max}$, and $\mu_{\max}-m$. Ellipticity and the two surface-brightness-related quantities show descriptive color gradients in parts of the projection, whereas Kron radius is more dispersed. These patterns describe the selected UMAP projection only; they do not establish that DINOv2 is necessary for the appearance of the projection or that UMAP preserves a global physical manifold.

\begin{figure*}[ht!]
\plotone{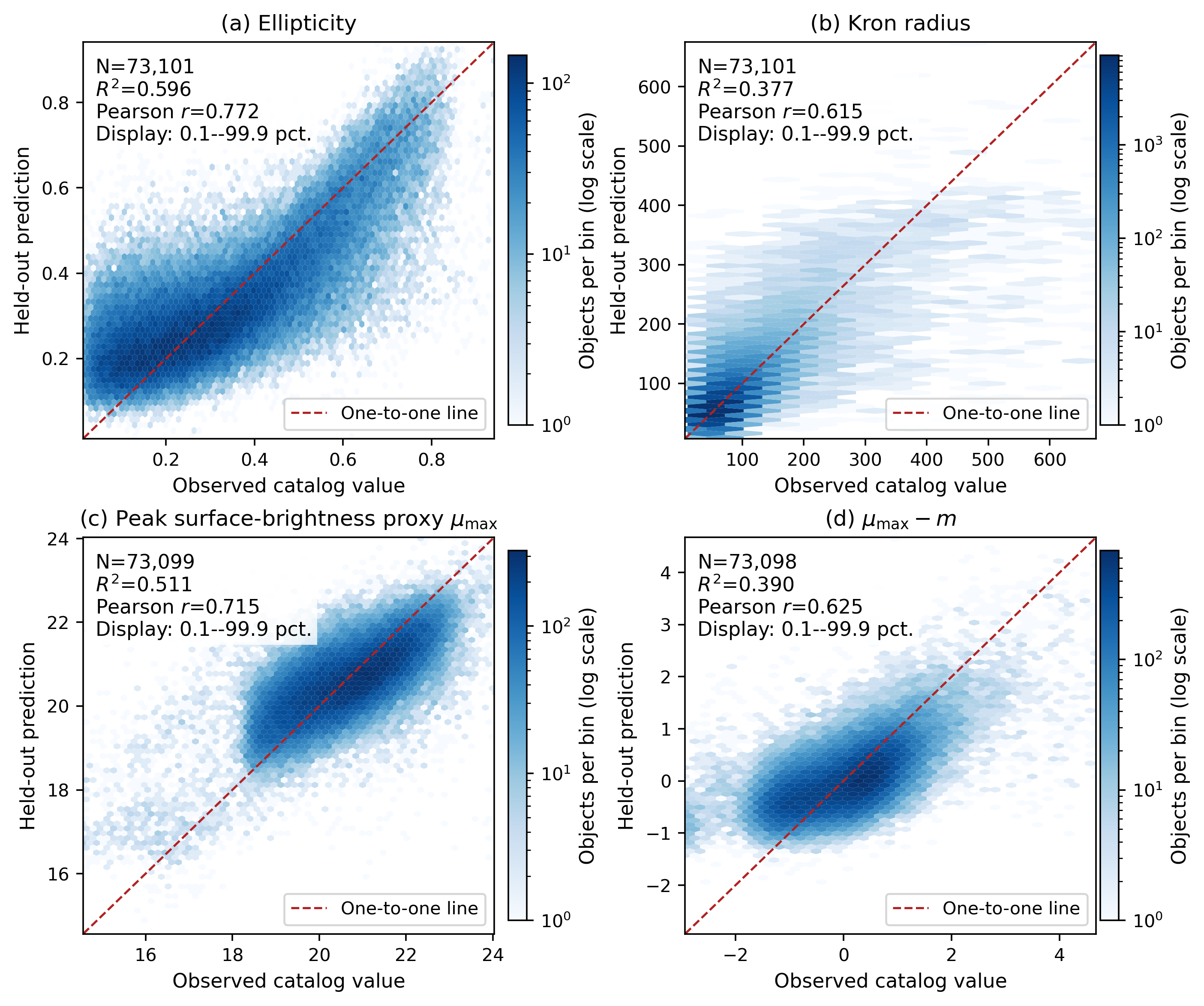}
\caption{Held-out ridge-regression probes of four catalog quantities from the fixed DINOv2 embedding. Hexagonal bins show test-object density on a logarithmic scale; the dashed line is the one-to-one relation. To prevent a small number of extreme catalog values from compressing the visible density, each panel displays the joint 0.1--99.9 percentile range of observed and predicted values; all reported metrics retain the complete held-out sample. Feature standardization and RidgeCV selection use the 80\% training partition only. Insets report the held-out sample size, $R^2$, and Pearson $r$; complete MAE, RMSE, and Spearman $\rho$ values are given in Table~\ref{tab:regression}.}
\label{fig:regression_probes}
\end{figure*}

We therefore quantify recoverable information directly in the original embedding with held-out ridge probes. Figure~\ref{fig:regression_probes} displays the density of observed and predicted values, and Table~\ref{tab:regression} gives the complete metrics. Ellipticity has $R^2=0.596$, MAE $=0.096$, RMSE $=0.125$, Pearson $r=0.772$, and Spearman $\rho=0.728$ on 73,101 test objects. The corresponding $R^2$ values are 0.377 for Kron radius, 0.511 for $\mu_{\max}$, and 0.390 for $\mu_{\max}-m$, with Pearson correlations of 0.615, 0.715, and 0.625, respectively. Kron radius and $\mu_{\max}-m$ show visibly broader departures from the one-to-one relation. Thus, the representation contains linearly recoverable information about each catalog quantity, but the strength and calibration of that information are target dependent.

\begin{deluxetable*}{llrrrrrrr}
\tabletypesize{\scriptsize}
\tablewidth{0pt}
\tablecaption{Held-out ridge probes for explicitly defined catalogue fields.\label{tab:regression}}
\tablehead{\colhead{Target} & \colhead{Unit} & \colhead{Train N} & \colhead{Test N} & \colhead{$R^2$} & \colhead{MAE} & \colhead{RMSE} & \colhead{Pearson $r$} & \colhead{Spearman $\rho$}}
\startdata
Ellipticity $1-B/A$ & dimensionless & 292,403 & 73,101 & 0.596 & 0.096 & 0.125 & 0.772 & 0.728 \\
Kron aperture major semi-axis & pixel & 292,403 & 73,101 & 0.377 & 19.914 & 54.297 & 0.615 & 0.656 \\
Peak surface brightness $\mu_{\max}$ & mag arcsec$^{-2}$ & 292,395 & 73,099 & 0.511 & 0.683 & 0.901 & 0.715 & 0.689 \\
$\mu_{\max}-\mathrm{MAG}_{\mathrm{AUTO}}$ & mag & 292,393 & 73,098 & 0.390 & 0.546 & 0.732 & 0.625 & 0.609 \\
\enddata
\tablecomments{Each target is filtered independently and assigned to an 80/20 object-ID split with \texttt{random\_state}=42. Feature standardization and RidgeCV selection use the training partition only; all reported metrics are computed once on held-out test objects.}
\end{deluxetable*}

\subsection{Few-label Classification} \label{subsec:4.2}

\begin{figure*}[ht!]
\plotone{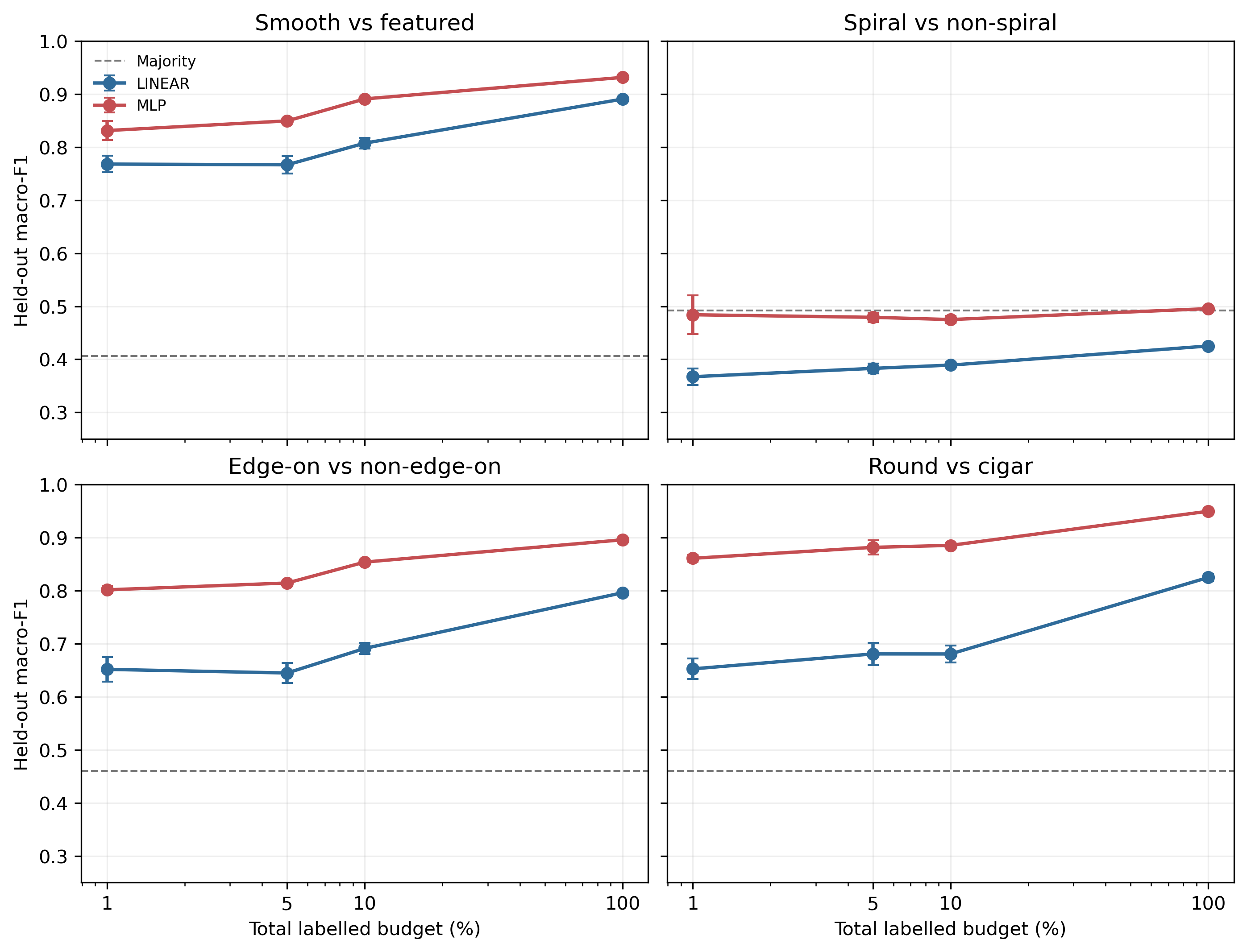}
\caption{Few-label classification with the frozen official DINOv2 embedding. Panels show four independently defined high-confidence Zoobot binary tasks. Lines give held-out macro-F1 for linear and MLP probes at total labelled budgets of 1\%, 5\%, 10\%, and 100\% of the task-specific development pool; each budget includes its training and validation subsets. Error bars are the standard deviation across seeds 42--44, and dashed lines show the task-specific majority baseline. Frozen and limited-fine-tuning results are kept separate; the latter are reported in Table~\ref{tab:fewlabel}.}
\label{fig:fewlabel_budget}
\end{figure*}

We evaluate four high-confidence binary tasks: smooth versus featured, spiral versus non-spiral, edge-on versus non-edge-on, and round versus cigar. Figure~\ref{fig:fewlabel_budget} reports frozen official-DINOv2 linear and MLP probes at total labelled budgets of 1\%, 5\%, 10\%, and 100\% of each task-specific development pool. Each point is the held-out macro-F1 mean over seeds 42--44, with error bars showing one standard deviation. At the one-percent budget, frozen MLP macro-F1 is $0.831\pm0.018$, $0.484\pm0.036$, $0.802\pm0.007$, and $0.861\pm0.006$ for smooth, spiral, edge-on, and round/cigar, respectively. The corresponding majority baselines are 0.406, 0.492, 0.461, and 0.460, and stratified-random baselines are approximately 0.50. Increasing the budget improves smooth, edge-on, and round/cigar to 0.931, 0.896, and 0.950 at 100\%, but spiral remains $0.496\pm0.001$. Thus, the spiral experiment does not establish useful discrimination under this highly imbalanced high-confidence contract.

Table~\ref{tab:fewlabel} reports test class composition, budget train/validation counts, trivial baselines, macro-F1, and per-class recall for the one-percent MLP comparison. Limited final-block fine-tuning gives $0.827\pm0.014$, $0.520\pm0.015$, $0.810\pm0.007$, and $0.850\pm0.028$ for smooth, spiral, edge-on, and round/cigar, respectively, and therefore does not uniformly improve over the frozen MLP. For spiral, limited fine-tuning attains negative-class recall of only 0.102 while positive-class recall is 0.961; its slightly higher macro-F1 does not resolve the class-collapse concern. The supported conclusion is task-specific recoverability for smooth, edge-on, and round/cigar under the stated Zoobot label contract, not a single performance claim across all tasks.

\begin{deluxetable*}{lrrrrrrrr}
\tabletypesize{\scriptsize}
\tablewidth{0pt}
\tablecaption{One-percent total-supervision-budget MLP results on task-specific held-out partitions.\label{tab:fewlabel}}
\tablehead{\colhead{Task} & \colhead{Test $N_-/N_+$} & \colhead{Budget train/val} & \colhead{Majority F1} & \colhead{Random F1} & \colhead{Frozen MLP F1} & \colhead{Frozen recall $-/+$} & \colhead{Limited MLP F1} & \colhead{Limited recall $-/+$}}
\startdata
Smooth vs featured & 6,043/13,105 & 612/154 & 0.406 & 0.497 & 0.831$\pm$0.018 & 0.815/0.867 & 0.827$\pm$0.014 & 0.768/0.888 \\
Spiral vs non-spiral & 314/9,794 & 324/81 & 0.492 & 0.501 & 0.484$\pm$0.036 & 0.273/0.821 & 0.520$\pm$0.015 & 0.102/0.961 \\
Edge-on vs non-edge-on & 18,653/3,186 & 698/175 & 0.461 & 0.500 & 0.802$\pm$0.007 & 0.899/0.805 & 0.810$\pm$0.007 & 0.958/0.631 \\
Round vs cigar & 1,563/8,987 & 337/85 & 0.460 & 0.498 & 0.861$\pm$0.006 & 0.904/0.923 & 0.850$\pm$0.028 & 0.733/0.960 \\
\enddata
\tablecomments{For each task, the one-percent budget is sampled from the development partition and then split 80/20 into training and validation; no additional labelled validation set is used. Values are mean $\pm$ standard deviation over seeds 42--44 on the same held-out test partition. The majority reference predicts the training-subset majority class for every test object; the stratified-random reference samples positives with the training-subset positive fraction and is averaged over the same seeds. ``Frozen'' trains only the MLP head; ``Limited'' uses one head-only warm-up epoch and then unfreezes the final transformer block, final normalization, and head for at most four epochs.}
\end{deluxetable*}

As external numerical context, AION-1 reports Galaxy Zoo 10 classification accuracies of 84.0\%, 87.2\%, and 86.5\% for its B, L, and XL frozen encoders, respectively, on approximately 8,000 Legacy Survey galaxies \citep{parker_aion_2025}. Under the same published benchmark, a frozen DINOv2 ViT-g/14 encoder with an MLP head achieves 71.4\%, an end-to-end EfficientNet-B3 achieves 80.0\%, and ZooBot achieves 89.6\%. These values place frozen astronomical image representations in a concrete external numerical context. They are reported separately from Table~\ref{tab:fewlabel} because the external benchmark is a ten-class Legacy Survey task evaluated with accuracy, whereas our experiments comprise four binary Euclid Q1 tasks evaluated with held-out macro-F1. AstroPT provides additional same-survey context, but its cohort, labels, split, model family, and metric aggregation also differ from the present evaluation, so these values are not used as a direct ranking.

\begin{figure*}[ht!]
\plotone{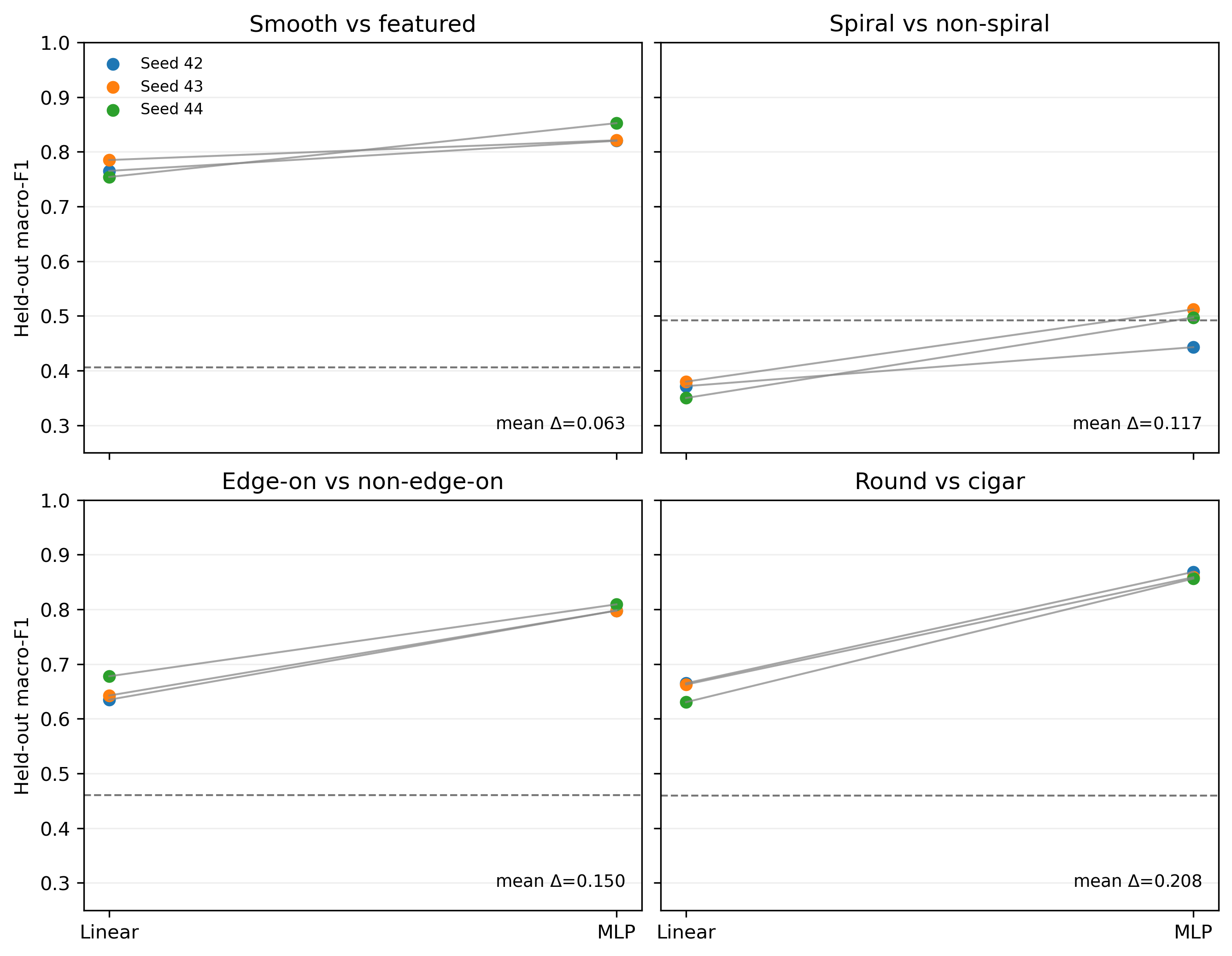}
\caption{Paired one-percent-total-budget results for frozen official-DINOv2 probes. Each line connects held-out macro-F1 for linear and MLP heads using the same task, seed, budget and test partition; colors identify seeds 42--44, dashed lines show majority baselines, and each panel reports the mean paired difference. Although every paired run favors the MLP head, the spiral values remain near trivial baselines. The comparison does not establish unconditional MLP superiority for other tasks, label regimes, or backbone states.}
\label{fig:fewlabel_paired_heads}
\end{figure*}

At the one-percent total labelled budget, Figure~\ref{fig:fewlabel_paired_heads} shows the seed-level paired comparison between linear and MLP heads under the same frozen official-DINOv2 contract. Across seeds 42--44, the MLP-minus-linear macro-F1 difference is 0.036--0.098 for smooth versus featured, 0.071--0.147 for spiral versus non-spiral, 0.131--0.163 for edge-on versus non-edge-on, and 0.196--0.226 for round versus cigar. All twelve paired runs favor the MLP relative to the corresponding linear head, but the spiral MLP remains at or below the trivial baselines and therefore does not demonstrate a useful task result. This task- and budget-specific comparison does not support an unconditional claim that an MLP is superior in other label regimes or backbone states.

Appendix~\ref{app:historical_ssl} preserves the submitted Euclid-SSL, official-DINOv2, and random-initialization training traces as historical optimization diagnostics. Because the configurations are not matched, these traces are not included in the quantitative initialization or representation comparison reported in the main text.

\subsection{Anomaly Detection} \label{subsec:4.3}

\begin{figure*}[ht!]
\plotone{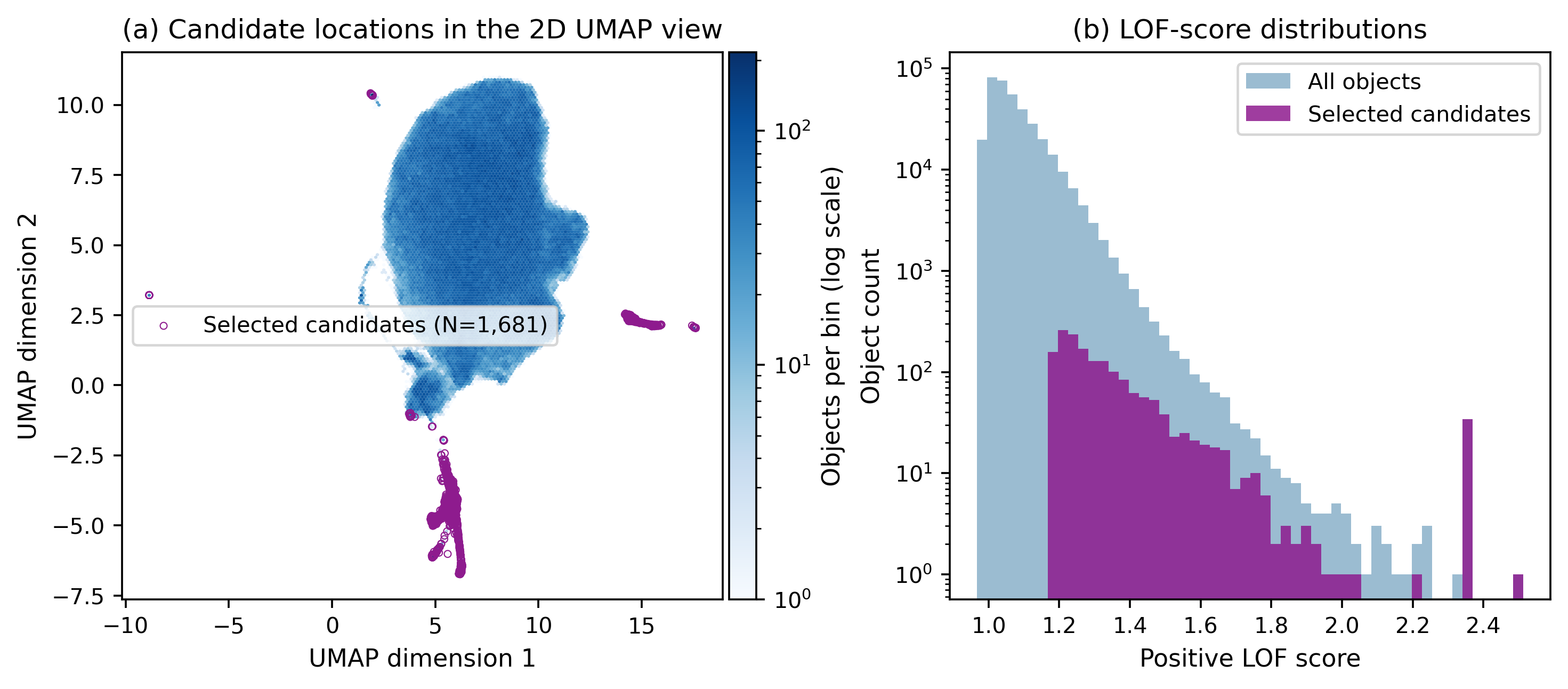}
\caption{Historical LOF--UMAP candidate selection. (a) Logarithmic object density in the same saved two-dimensional UMAP projection described in Section~\ref{subsec:3.2}, with the same 1,681 candidates from the original intersection rule overplotted. The projection uses $n_{\mathrm{neighbors}}=15$, $\mathrm{min\_dist}=0.1$, the Euclidean metric, and \texttt{random\_state}=42. (b) Positive LOF-score distributions for all objects and selected candidates on a logarithmic count axis. LOF used the raw 384-dimensional embeddings with $k=100$ and \texttt{contamination}=0.1; the secondary filter retained objects beyond the 98th percentile of centroid distance in that saved projection. The visualization documents a configuration-dependent quality-control candidate list, not a complete astrophysical anomaly census.}
\label{fig:anomaly_selection}
\end{figure*}

We apply the historical two-stage procedure described in Section~\ref{subsec:3.4} to prioritize unusual or problematic images. Figure~\ref{fig:anomaly_selection} shows the resulting 1,681 candidates in the two-dimensional UMAP view and compares their positive LOF-score distribution with that of the full embedding sample. The candidates concentrate in several peripheral regions of this particular projection and have a higher-score distribution than the parent sample. Because selection combines LOF on the raw 384-dimensional embedding with a 98th-percentile UMAP-centroid-distance cut, both the count and membership depend on $k$, contamination, normalization, UMAP settings, and the intersection rule.

\begin{figure*}[ht!]
\plotone{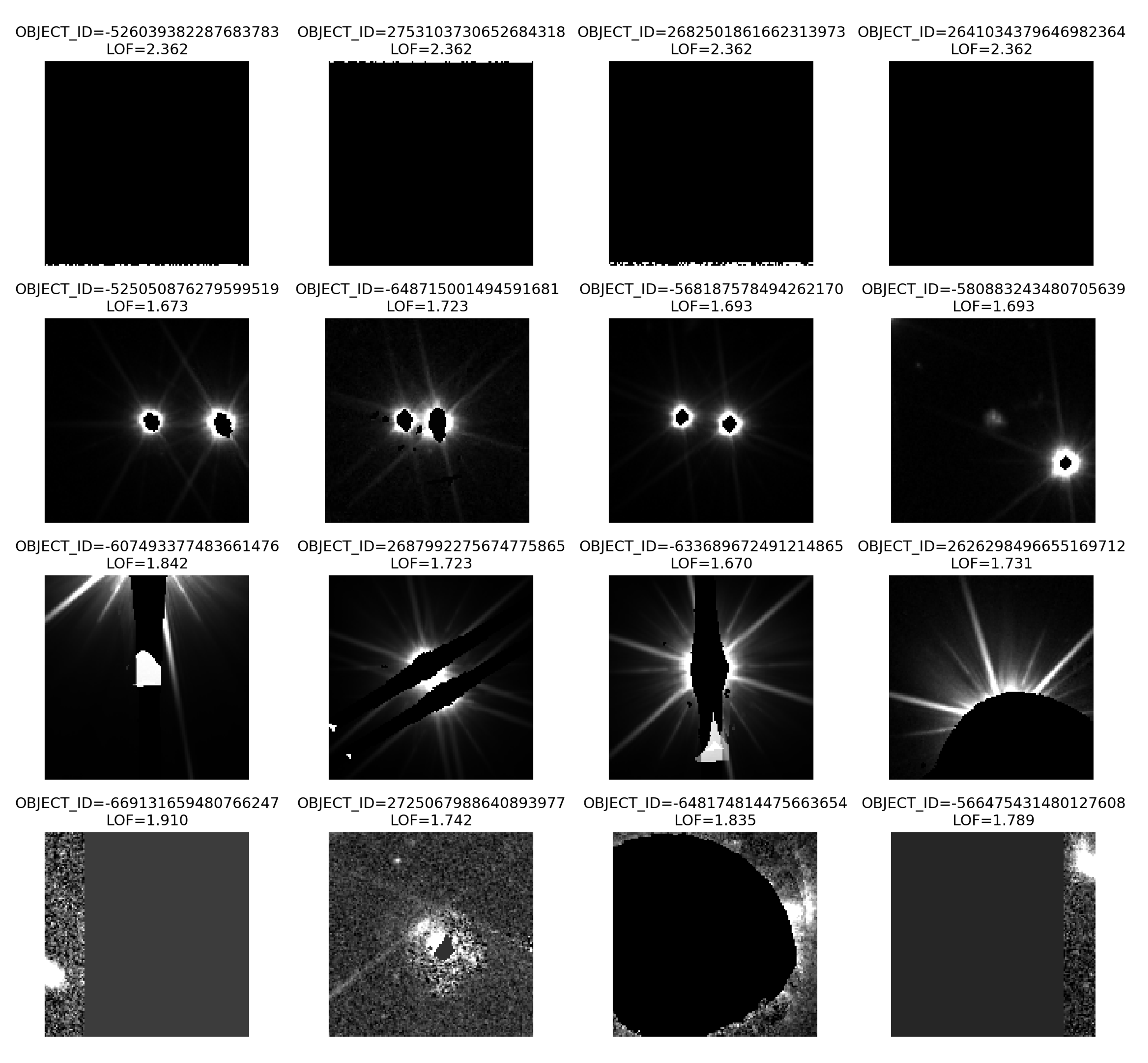}
\caption{Representative high-ranked candidates from the historical anomaly workflow. The examples are dominated by blank or truncated cutouts, saturation and diffraction structure, blends, and low-signal images; they therefore demonstrate the workflow's immediate value for data-quality control rather than confirming astrophysical anomalies.}
\label{fig:anomaly_examples}
\end{figure*}

Figure~\ref{fig:anomaly_examples} presents illustrative examples selected from the high-ranked historical candidate set. The original analysis did not retain a reproducible 100-object selection list, rater record, or category-count protocol, so these panels cannot support prevalence estimates. The displayed examples illustrate four types of output:

First, the displayed candidates include imaging artifacts or corrupted cutouts, such as nearly blank images, partially missing data, and edge-truncated sources plausibly associated with cutout boundaries or data-ingestion issues.

Second, some displayed candidates are dominated by bright foreground stars or saturated point-spread-function features. These objects exhibit diffraction spikes and central saturation that differ from typical extended galaxy profiles. Their selection demonstrates sensitivity to image-quality failures, but it does not establish a calibrated detection rate.

Third, the examples include blended or potentially interacting systems, where multiple sources overlap within the same cutout. These images may show double nuclei, irregular light distributions, or asymmetric structures. Their interpretation remains provisional because the present visual inspection does not provide independent astrophysical labels.

Finally, the examples include low-signal or noise-dominated images in which the target is faint or embedded in strong background fluctuations. These examples support quality-control triage as a plausible use of the candidate ranking.

The visual examples show that the ranking can surface blank, truncated, saturated, diffraction-dominated, blended, or noise-dominated cutouts. We therefore interpret the demonstrated workflow as a qualitative quality-control and inspection-prioritization example. Establishing category prevalence or an astrophysical anomaly catalogue would require a preregistered selection rule, retained object list, explicit rater protocol, sensitivity analysis, and independent validation.

\subsection{Similarity Search} \label{subsec:4.4}

\begin{figure*}[ht!]
\plotone{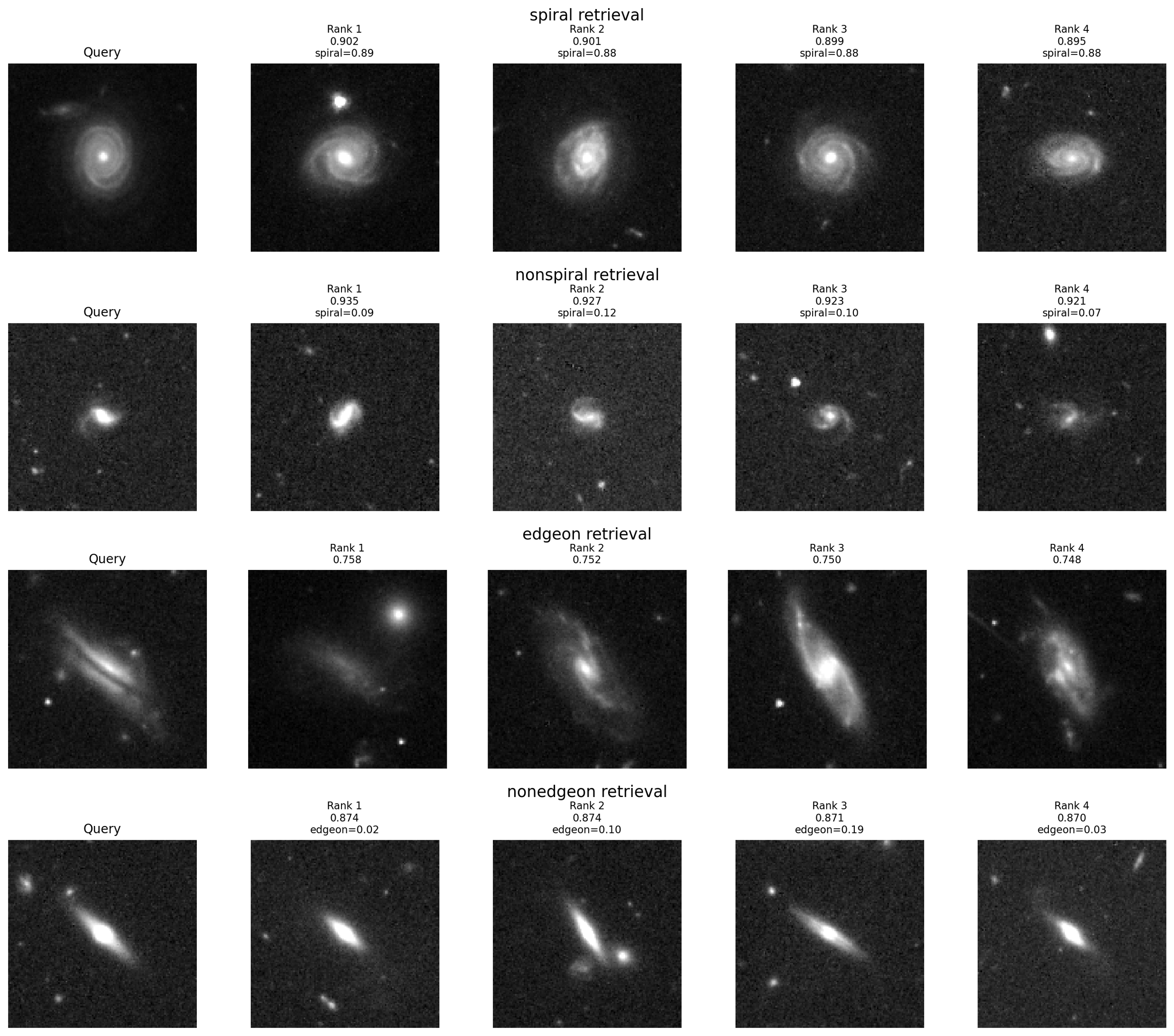}
\caption{Qualitative retrieval examples for spiral/non-spiral and edge-on/non-edge-on queries. Each row begins with the query and shows four ranked neighbors. The first displayed number is the manually weighted combined retrieval score; the second, when present, is the corresponding Galaxy Zoo morphology probability and was not available for every object. The historical weights were heuristic and were not fitted on an independent validation-query set.}
\label{fig:retrieval_group_one}
\end{figure*}

\begin{figure*}[ht!]
\plotone{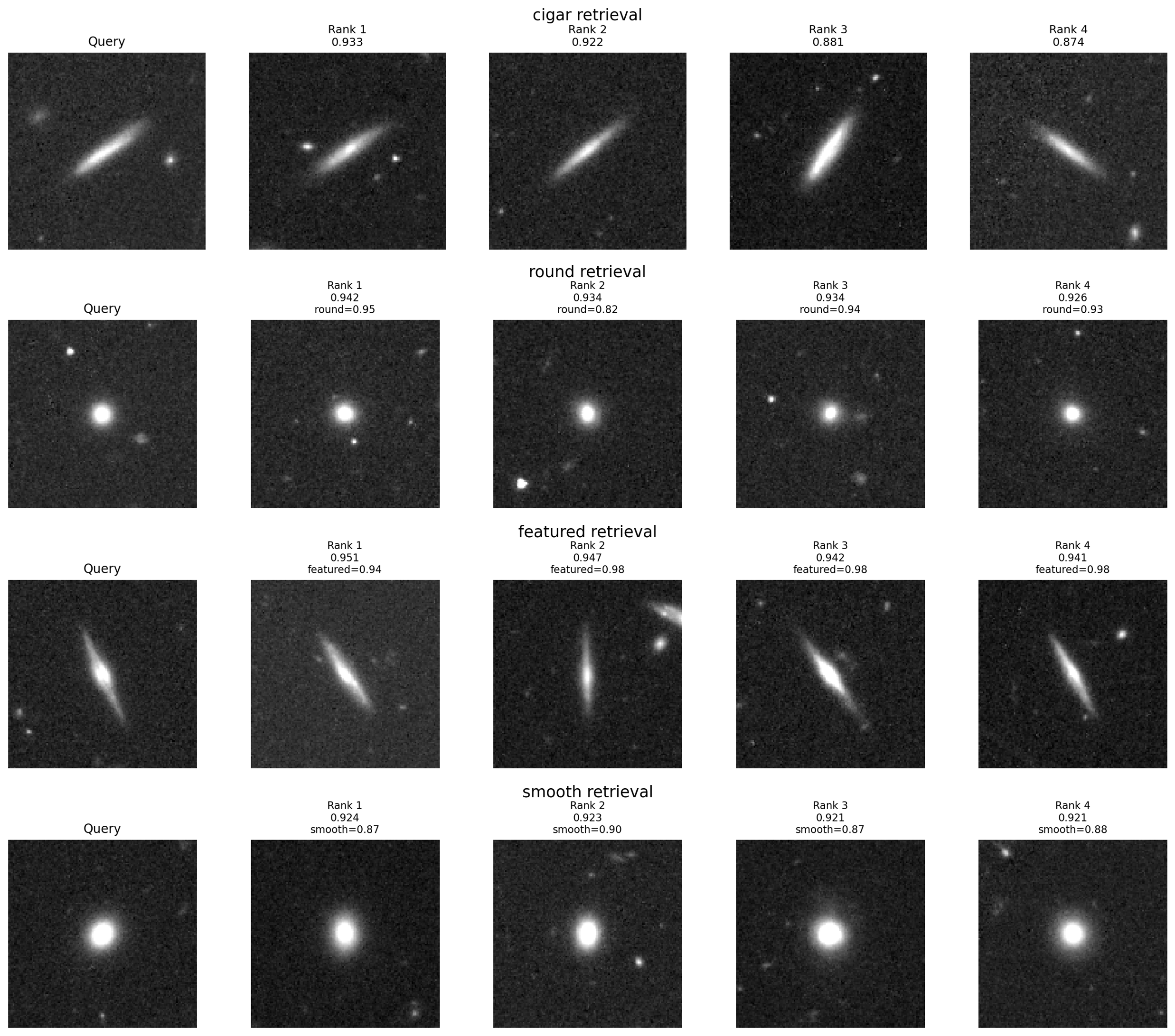}
\caption{Qualitative retrieval examples for cigar, round, featured, and smooth queries. Each row begins with the query and shows four ranked neighbors. The first displayed number is the manually weighted combined retrieval score; the second, when present, is the corresponding Galaxy Zoo morphology probability. These panels illustrate retrieval behavior but do not constitute a calibrated performance benchmark.}
\label{fig:retrieval_group_two}
\end{figure*}

Beyond classification and anomaly detection, the representation supports similarity-based exploration. The historical examples in Figures~\ref{fig:retrieval_group_one} and~\ref{fig:retrieval_group_two} first identify candidates by embedding cosine similarity and then apply the manually weighted reranking described in Section~\ref{subsec:3.5}. For each query, four neighbors are displayed with the combined score and, where available, a morphology-label probability.

In the displayed examples, several spiral queries retrieve neighbors with visible spiral structure, while several non-spiral, edge-on, elongated, and round queries retrieve images with qualitatively similar appearance. These selected panels illustrate plausible retrieval behavior but do not estimate task-level retrieval accuracy or robustness.

Where Galaxy Zoo probabilities are available, several returned neighbors agree with the query morphology. This observation is descriptive rather than an independent validation because label-related information contributed to the historical reranking score and the displayed queries were not selected through a preregistered evaluation protocol.

In some panels, retrieved images are not accompanied by explicit label annotations because the search covers sources outside the high-confidence Galaxy Zoo subset. Their apparent visual resemblance to the query is anecdotal in the present analysis and cannot be used to claim generalization beyond the labeled subset without independent annotations.

The displayed combined scores generally decrease smoothly with rank, as expected from sorting by the historical reranking rule. This behavior is not a calibration test and does not establish that score differences correspond to fixed changes in morphological similarity.

To quantify the limited retrieval check, we evaluated four fixed held-out queries per task. At $k=10$, mean label consistency for embedding-only, 0.7/0.3, and 0.5/0.5 embedding/physical weighting was 0.500/0.950/0.975 for edge-on versus non-edge-on, 0.825/0.925/0.975 for round versus cigar, 0.875/0.850/0.925 for smooth versus featured, and 0.500/0.525/0.550 for spiral versus non-spiral. The equal-weight ablation had the highest observed mean in this small audit, but four queries per task and the query-level scatter are insufficient to select a generally optimal weight. Because this is a two-component ablation rather than a direct validation of the four-component historical weights, Figures~\ref{fig:retrieval_group_one} and~\ref{fig:retrieval_group_two} remain qualitative demonstrations.

Taken together, these examples show that the embedding can seed an interpretable retrieval workflow, while quantitative claims depend on a separately defined query set, ranking rule, and evaluation metric. The present panels are therefore retained as qualitative demonstrations rather than evidence of fully unsupervised or generally accurate retrieval.

\section{Discussion} \label{sec:discussion}

The results presented in Sections~\ref{sec:4} show that the fixed representation contains linearly recoverable information about the evaluated catalogue quantities and supports task-dependent downstream probes. In this section, we place these findings in a broader context, clarify the role of representation learning within an AI-ready workflow, and discuss both the implications and limitations of the current approach.

The present results should be interpreted relative to established representation-learning and Euclid Q1 products. Large-sample self-supervised representations already support similarity search and lightweight linear classification \citep{stein_similarity_2021,stein_mining_2022}, and supervised galaxy representations can transfer across multiple morphology tasks \citep{walmsley_practical_2022}. On Euclid Q1 itself, Euclid Collaboration: Siudek et al. \citeyearpar{2025arXiv250315312E} trained the AstroPT multimodal model and evaluated morphology, redshift, similarity-search, and outlier tasks, while Euclid Collaboration: Walmsley et al. \citeyearpar{2025arXiv250315310E} provided the detailed morphology catalogue that underlies our high-consensus labels. AION-1 demonstrates a broader frozen-encoder programme across several surveys and modalities \citep{parker_aion_2025}. These works establish the task and model landscape. Because the inputs, labels, splits, modalities, and metrics are not matched, we use them as external context rather than a direct performance ranking; within the declared Euclid VIS contract, our contribution is a complementary, auditable cutout-to-embedding and downstream-evaluation interface.

Against that background, the contribution supported here is an auditable implementation layer rather than a new task or foundation-model claim. The released cutout service standardizes Euclid VIS inputs, persistently reuses matching cutout products to avoid redundant extraction, the official-DINOv2 interface exposes a fixed 384-dimensional representation without additional pretraining, and the downstream scripts reuse that interface for regression, classification, candidate prioritization, and retrieval. The fixed object-ID splits, complete metric tables, seed-level comparisons, and explicit separation of quantitative from qualitative evidence make the limits of each demonstration inspectable. This complements AstroPT's Euclid-specific multimodal learning and the Zoobot morphology catalogue; it does not replace either product or claim priority for their scientific capabilities.

A central outcome of the controlled benchmark is that probe performance depends strongly on the task definition. The paired comparison in Figure~\ref{fig:fewlabel_paired_heads} shows a nonlinear-head improvement over the corresponding linear probe in all twelve frozen one-percent runs, but that relative gain does not make the imbalanced spiral task useful because its absolute result remains near trivial baselines. Head architecture and scientific task validity must therefore be assessed separately.

The retained historical SSL runs, documented in Appendix~\ref{app:historical_ssl}, show why future domain-adaptation experiments should predefine matched architectures, inputs, labels, splits, seeds, checkpoint-selection rules, and evaluation tasks before comparative claims are made.

From a methodological perspective, the key contribution of this work lies in the integration of three components: (1) a standardized cutout pipeline built on the NADC platform, (2) a pretrained representation model, and (3) a common embedding interface for several downstream analyses. Figure~\ref{fig:fewlabel_budget} shows one-percent-budget frozen MLP values of 0.802--0.861 for smooth, edge-on, and round/cigar, while spiral is 0.484 and unsupported by the trivial baselines. Limited final-block fine-tuning is 0.810--0.850 for the three supported tasks and 0.520 for spiral, with no uniform improvement over frozen probes. Figures~\ref{fig:anomaly_selection} and~\ref{fig:anomaly_examples} illustrate qualitative quality-control triage, and Figures~\ref{fig:retrieval_group_one} and~\ref{fig:retrieval_group_two} provide qualitative retrieval examples. These results support reuse of the same representation, but they also show that performance and scientific interpretation must be stated separately for each task.

Despite these strengths, several limitations should be acknowledged. First, DINOv2 is not specifically optimized for astronomical imaging, and the domain gap may limit sensitivity to subtle features. Second, the labels are high-confidence Zoobot predictions under a conditional question tree, not an independent human-labelled benchmark; the resulting filters favor unambiguous cases and the spiral cohort is strongly imbalanced. The spiral task therefore illustrates a limitation of this embedding under a highly imbalanced label contract. Third, the anomaly workflow depends on the LOF neighborhood size, contamination setting, raw-embedding scale, UMAP hyperparameters, and a global-distance statistic that UMAP is not designed to preserve. The 1,681 candidates are therefore configuration dependent, and the retained examples do not provide a reproducible artifact-prevalence estimate. Automated quality assessment has also been studied directly for astronomical image collections \citep{2020AJ....159..170T}; in our case, however, the LOF ranking is not a trained quality classifier, so quality control remains a qualitative use case rather than a calibrated label.

The historical Euclid-specific SSL records remain unsuitable for a matched comparison with official DINOv2 because their architecture, inputs, feature dimensions, and retained split contract differ. Future work should predefine the architecture, labels, splits, seeds, checkpoint-selection rule, and evaluation tasks before comparing domain-adapted and generic pretrained representations. Cross-survey morphology experiments show that changes in resolution and noise can produce measurable data shift and may require explicit adaptation \citep{ye_galaxy_2025}; transfer beyond the present Euclid VIS sample must therefore be revalidated rather than assumed. Possible extensions include multi-band inputs and physically motivated augmentations. General self-supervised multimodal learning and astronomy-specific fusion reviews identify modality alignment, fusion strategy, and heterogeneous data quality as explicit design choices \citep{zong_self-supervised_2025,shao_deep_2026}; these extensions require a new controlled experiment rather than reinterpretation of the current diagnostic.

In summary, this work demonstrates a scalable AI-ready interface for several analyses of the released Euclid Q1 cutout sample. In this paper, scalability is used operationally: the service completed production at the 365,513-cutout release scale, supports batch-oriented handling of independent requests, and retains generated products for subsequent matching requests. The approximately 48-hour initial production record corresponds to an estimated rate of $2.12$ cutouts s$^{-1}$, while persistent product reuse avoids repeating mosaic extraction during later access. These complementary stages support a system-level workflow rather than relying on raw per-cutout speed alone. Downstream scientific performance remains task dependent.

\section{Data and Code Availability}\label{sec:Data and Code Availability}

The reviewed implementation of the Euclid VIS AI-ready pipeline is publicly available in the GitHub repository \url{https://github.com/xiejhhhhhh/Making-Euclid-VIS-Imaging-AI-Ready}. The exact software version used for this revision is release \texttt{v1.1.0} (\url{https://github.com/xiejhhhhhh/Making-Euclid-VIS-Imaging-AI-Ready/releases/tag/v1.1.0}), distributed under the MIT License and archived on Zenodo \dataset[doi:10.5281/zenodo.21629100]{https://doi.org/10.5281/zenodo.21629100} \citep{xie_euclid_ai_ready_software_2026}. The release contains the revised question-tree label and total-supervision-budget contracts, canonical scripts, fixed environment information, scientific contract tests, compact result summaries, and code-to-paper mappings used to audit the revised analyses.

The pipeline operates on Euclid cutout images supplied in FITS format. To obtain these cutouts, users may use the Euclid Image Cutout Service, which automates batch extraction of galaxy-centred VIS stamps from survey mosaics hosted by the NADC and is released separately at \url{https://github.com/xiejhhhhhh/Euclid-Image-Cutout-Service}. The analysis reads the released $128\times128$ arrays and applies the exact $128\rightarrow224$ pixel-to-tensor transform described in Section~\ref{subsec:2.2}.

The MIT License applies only to the author-maintained software release. The Galaxy Zoo Euclid Q1 catalogue is distributed separately under CC BY 4.0 at \dataset[doi:10.5281/zenodo.15106473]{https://doi.org/10.5281/zenodo.15106473}; the official DINOv2 implementation and checkpoint are obtained under the upstream Apache 2.0 terms; and Euclid/NADC images and cutouts remain subject to their originating data-access and reuse policies rather than the software MIT License. The Euclid VIS cutout dataset used here is available via the NADC at \url{https://nadc.china-vo.org/res/r101833/}. Euclid images, source catalogues, pretrained checkpoints, object-level embeddings, and private infrastructure logs are not redistributed in the software archive; their access boundaries and the inputs required by each canonical runner are documented in the release.

\section{Conclusion} \label{sec:conclusion}

In this work, we transform released Euclid VIS cutouts into standardized model inputs and use a pretrained DINOv2 vision transformer to provide a common 384-dimensional feature interface for several downstream analyses. Descriptive UMAP projections show local gradients, while held-out regression and classification quantify which catalog and morphology information is recoverable from the released embedding.

At the infrastructure level, the cutout service has operated at the 365,513-cutout release scale and supports batch-oriented processing and repeated access through persistent product reuse.

The same representation supports regression, few-label classification, candidate prioritization, and similarity-based exploration, but the evidence is not uniform across tasks. Frozen one-percent-budget MLP performance is 0.802--0.861 for three supported tasks, while the strongly imbalanced spiral task remains near trivial baselines. The 1,681 anomaly candidates are configuration dependent and the retained examples are illustrative rather than a prevalence audit; the retrieval examples remain qualitative because their historical reranking weights were manually specified.

Looking forward, the framework presented here provides a practical pathway for integrating machine learning into the analysis of large survey datasets. The embedding space can be extended to incorporate multi-band observations, spectroscopic information, or time-domain variability, enabling richer scientific interpretations. At the same time, the challenges encountered in domain-specific self-supervised learning point to the need for future work on physically informed training strategies and domain-adapted foundation models.

Taken together, this study illustrates how a carefully constructed workflow—linking data infrastructure, representation learning, and downstream analysis—can transform large astronomical datasets into flexible and interpretable AI-ready resources for scientific discovery.

\appendix
\section{Historical Euclid-specific SSL Optimization Diagnostics}
\label{app:historical_ssl}

This appendix preserves historical optimization records that are useful for documenting a prior Euclid-specific self-supervised learning (SSL) attempt but are not part of the matched performance benchmark. The records should not be interpreted as evidence that official DINOv2 is generally superior to domain-specific SSL, or that the Euclid-specific representation collapsed.

The Euclid-specific SSL checkpoint was produced in this project with an eight-epoch DINO teacher--student objective on the released Euclid VIS cutouts, using a local small-ViT implementation with patch size 16, single-channel inputs, and 256-dimensional features. The checkpoint used a 5\% validation split. The retained downstream diagnostic traces use the spiral-versus-non-spiral task at a 1\% label fraction, seeds 42--44, 20 fine-tuning epochs, and five head-only warm-up epochs followed by joint optimization of the head and backbone. The official DINOv2 traces use ViT-S/14 with replicated three-channel inputs and 384-dimensional features, while the random-initialization traces use the local small-ViT configuration. These differences in architecture, input channels, feature dimension, and retained split records prevent a same-contract initialization comparison.

\begin{figure*}[ht!]
\plotone{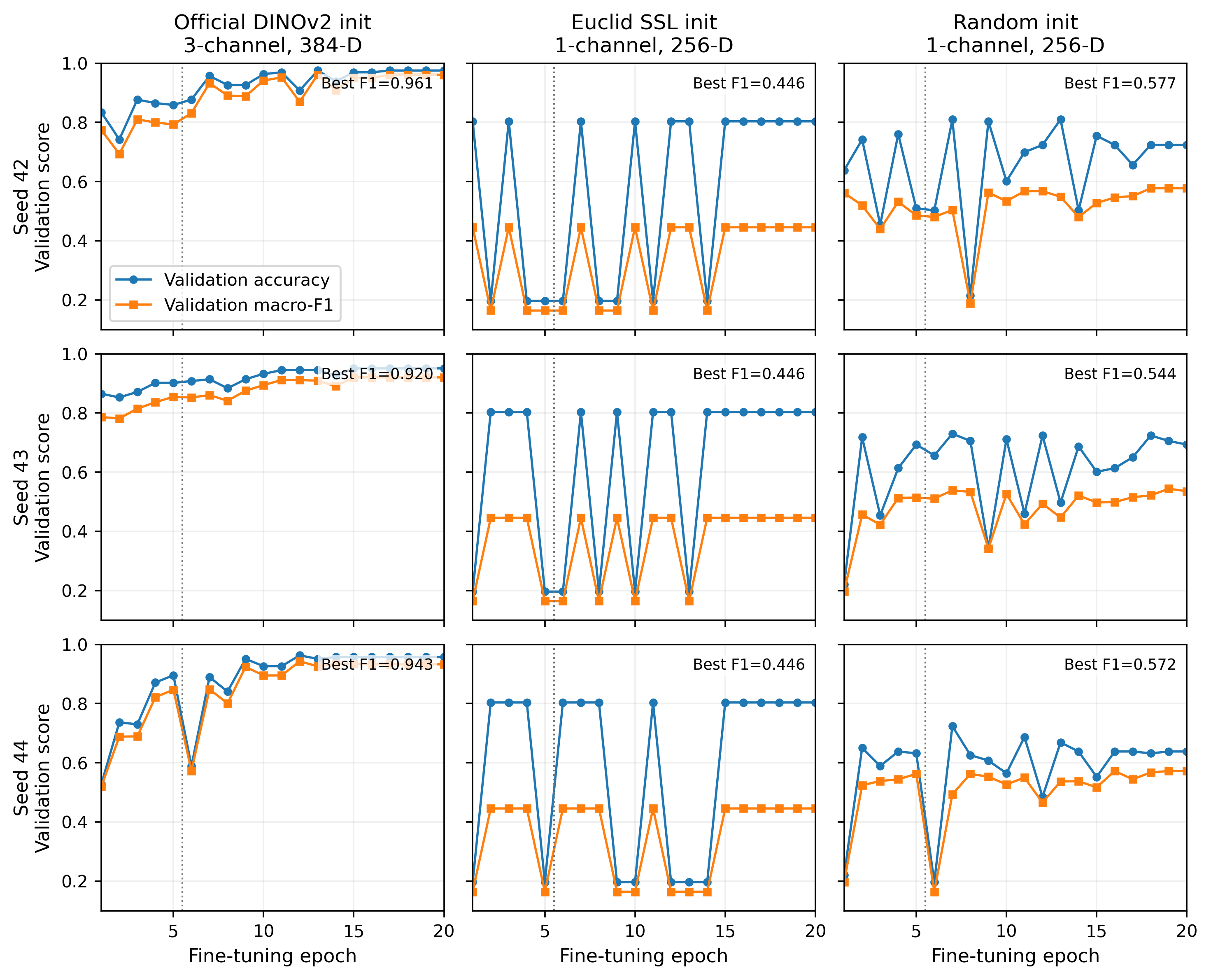}
\caption{Historical optimization diagnostics for unmatched official-DINOv2, Euclid-specific SSL, and random-initialization runs on the spiral-versus-non-spiral task. Columns show the three initialization records and rows show seeds 42--44. Curves report validation accuracy and macro-F1 over 20 epochs, vertical dotted lines mark the transition after five head-only warm-up epochs, and annotations give the best macro-F1 retained for each run. The original bottom-center \texttt{frac100} panel has been replaced by the corresponding saved \texttt{frac001}, seed-44 history. Official DINOv2 uses replicated three-channel inputs and 384-dimensional features, whereas the Euclid-SSL and random runs use single-channel inputs and 256-dimensional features. The panels are historical training diagnostics, not a matched initialization benchmark, and they do not establish general model superiority or representation collapse.}
\label{fig:historical_ssl}
\end{figure*}

Figure~\ref{fig:historical_ssl} summarizes the retained validation-accuracy and macro-F1 trajectories for the three unmatched initialization records across seeds 42--44. The best validation macro-F1 values in these traces are 0.961, 0.920, and 0.943 for official DINOv2; 0.446 for each retained Euclid-SSL run; and 0.577, 0.544, and 0.572 for random initialization. These values and curve trajectories describe the saved runs only. Because the three columns do not share the same architecture, input-channel, feature-dimension, or retained split contract, Figure~\ref{fig:historical_ssl} is used for provenance and optimization diagnosis rather than model ranking, representation-collapse diagnosis, or a general claim about pretraining versus domain-specific SSL.

Future domain-adaptation comparisons should predefine matched architectures, inputs, labels, object-ID splits, seeds, checkpoint-selection rules, and evaluation tasks before making comparative claims. The separate 9,729-image RTX 4070 resource estimate remains an internal audit record and is not used here as evidence for the released cutout throughput or for this SSL comparison.

\clearpage

\section*{Acknowledgements}

This work was supported by the Strategic Priority Research Program of the Chinese Academy of Sciences (XDB0550101), and the National Natural Science Foundation of China (NSFC; 12403102, 12373110, 12273077). Computing resources were provided by the National Astronomical Observatories, Chinese Academy of Sciences. Data resources are supported by the China National Astronomical Data Center, the CAS Astronomical Data Center, and the Chinese Virtual Observatory (China-VO). This work is also supported by the Astronomical Big Data Joint Research Center, co-founded by the National Astronomical Observatories, Chinese Academy of Sciences and Alibaba Cloud. This work makes use of data from the Euclid mission, and we acknowledge the Euclid Collaboration for providing the data used in this study. This work also makes use of the Galaxy Zoo Euclid (Q1) morphology catalogue, and we thank the Galaxy Zoo team and the many volunteers whose contributions made this work possible.

\section*{Author Contributions}

Jinhui Xie contributed to data acquisition, code development, image generation, and manuscript writing and revision. 
Yunfei Xu contributed to project supervision, data acquisition, funding acquisition, computational resource support, and manuscript review. 
Zhen Zhang contributed to data acquisition and code development. 
Lang Chen contributed to data acquisition. 
Chenzhou Cui contributed to project supervision, funding acquisition, and computational resource support.




\bibliography{sample701}{}
\bibliographystyle{aasjournalv7}



\end{document}